\documentclass[12pt,preprint]{aastex}

\begin{document}

\title{Hot White Dwarfs}

\author{Edward M.\ Sion}
\affil{Department of Astronomy \& Astrophysics, Villanova University}

\begin{abstract}
The article covers the physical properties and evolution of single white  dwarfs ranging in temperature from 20,000K to 200,000 and higher, the  hottest know electron-degenerate stars. After discussing the classification  of their spectra, the author reviews the known properties, parameters,  evolutionary state, as well as persisting and new puzzles regarding all  spectroscopic subclasses of Hot White Dwarfs: the hot DA white dwarfs, the  DAO white dwarfs, the PG1159 degenerates, the DO white dwarfs, the DB white  dwarfs, the DBA white dwarfs, and the Hot DQ white dwarfs (an entirely new  class). The most recent observational and theoretical advances are brought to bear on the topic.
\end{abstract}

\section{Introduction}

Research on hot white dwarfs during the past thirty years has greatly expanded, as many new discoveries, and the new questions they raise, have emerged from increasingly larger, deeper surveys conducted with multi-meter class ground-based telescopes, the {\it International Ultraviolet Explorer} ({\it IUE\/}), the {\it Hubble Space Telescope} ({\it HST\/}), the {\it Extreme Ultraviolet Explorer} ({\it EUVE\/}),  and the {\it Far Ultraviolet Spectroscopic Explorer} ({\it FUSE\/}).  This review will focus on white dwarfs ranging in temperature from 20,000~K up to 200,000~K and higher, which are the hottest white dwarf stars known.  Since the mid-20th century, the earliest spectroscopic surveys of white dwarf candidates from the proper motion selected samples of Willem Luyten and Henry Giclas were carried out by Jesse Greenstein, Olin Eggen, James Liebert, Richard Green, and others.  The selection criteria employed in many of these surveys did not reveal a large number of hot white dwarfs because the surveys lacked ultraviolet sensitivity and also missed objects with low flux levels in the optical.  Nevertheless, the earliest surveys quickly revealed that white dwarfs divide into two basic composition groups, with hydrogen-rich (the DA stars) and helium-rich atmospheric compositions (the DB and other non-DA stars).  The origin of this dichotomy still represents a major unsolved problem in stellar evolution, although theoretical advances in late stellar evolution made starting in the 1980s, as well as advances in modeling envelope physical processes and mass loss, have shed important new light on this puzzle \citep*{ch1-fontaine79,ch1-vvg79,ch1-iben83,ch1-schoen83,ch1-iben84,ch1-chayer95,ch1-unglaub98,ch1-unglaub00}.

The spectroscopic properties of white dwarfs are determined by a host of physical processes which control and/or modify the flow of elements and, hence, surface abundances in high gravity atmospheres:\ convective dredge-up, mixing and dilution, accretion of gas and dust from the interstellar medium and debris disks, gravitational and thermal diffusion, radiative forces, mass loss due to wind outflow and episodic mass ejection, late nuclear shell burning and late thermal pulses, rotation, magnetic fields, and possible composition relics of prior pre-white dwarf evolutionary states. Virtually all of these processes and factors may operate in hot white dwarfs, leading to the wide variety of observed spectroscopic phenomena and spectral evolution.

The basic thrust of research on hot white dwarfs is three-fold:\ (1) to elucidate the evolutionary links between the white dwarfs and their pre-white dwarf progenitors, whether from the asymptotic giant branch (AGB), the extended horizontal branch, stellar mergers, or binary evolution; (2) to understand the physics of the different envelope processes operating in hot white dwarfs as they cool; and (3) to disentangle and elucidate the relationships between the different spectroscopic subclasses and hybrid subclasses of hot white dwarfs as spectral evolution proceeds. This includes the source of photospheric metals, the chemical species observed, and the measured  surface abundances in hot degenerates. The evolutionary significance of certain observed ion species is complicated by the role of radiative forces and weak winds in levitating and ejecting elements at temperatures $>$20,000~K.

\section{Remarks on the Spectroscopic Classification of Hot White Dwarfs}

Before discussing the hot white dwarfs, a brief discussion of their spectroscopic classification is appropriate. The non-DA stars fall into six subclasses, including the PG\,1159 stars (75,000~K $< T_{\rm eff} <$ 200,000~K); DO (45,000~K $< T_{\rm eff} <$ 120,000~K), and DB (12,000~K $< T_{\rm eff} <$ 45,000~K).  The remaining three subclasses, which are too cool ($T_{\rm eff} <$ 12,000~K) to show helium, are DC (pure continuum, no lines), DQ (helium-dominated but with carbon molecular bands and/or atomic carbon dominating the optical spectrum), and DZ (helium dominated with lines of accreted metals and sometimes H).  The three cool spectral subgroups, as well as the PG\,1159 stars, are discussed elsewhere in this volume (see the chapters by M.\ Kilic and P.\ Dufour), and are not covered here.  For convenience, Table~\ref{t:ch1-symbols} lists the various primary white dwarf spectroscopic classification symbols, any of which can also be assigned as a secondary symbol to form hybrid spectral classes.  For example, the identified classes of hot, hybrid composition degenerates are predominantly DBA, DAB, DAO, and DOA. 

\begin{figure}
\figurenum{1}
\plotone{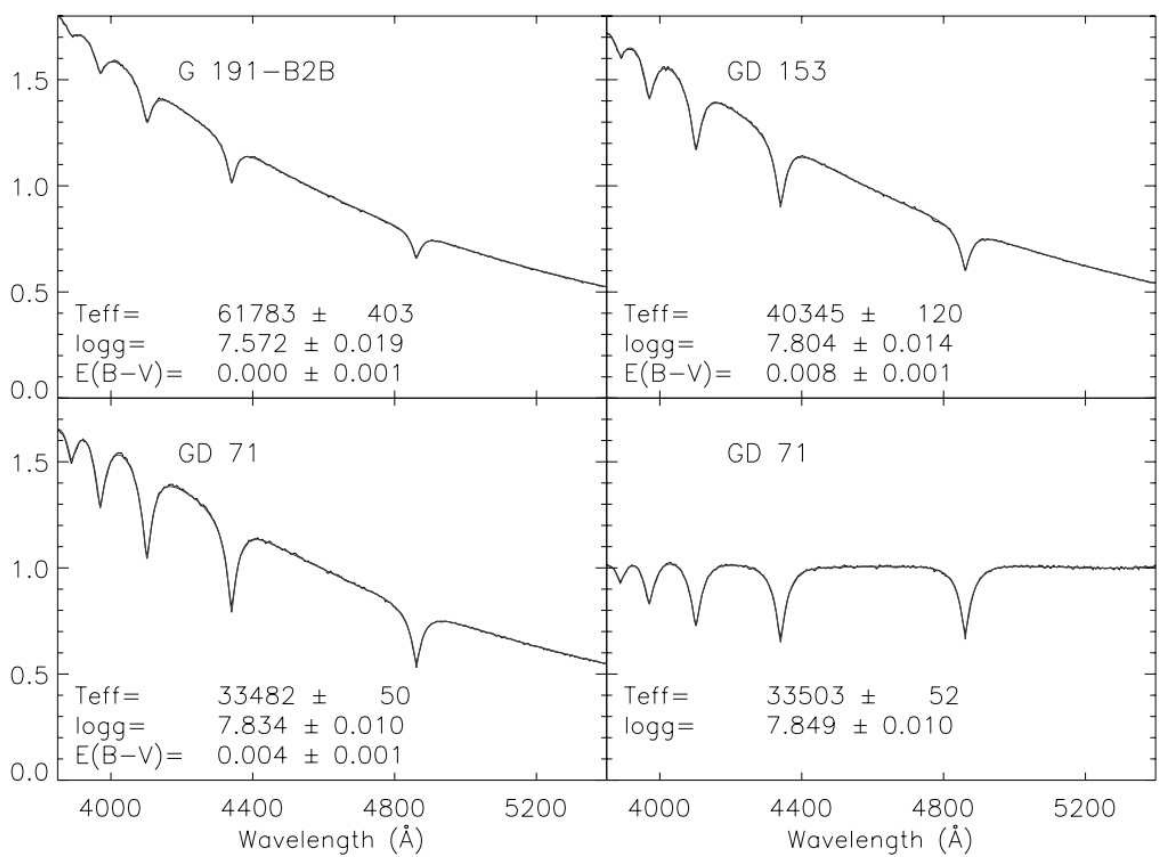}
\caption{
Spectra and model fits of three hot DA white dwarfs (G191-B2B, GD\,53, and GD\,71) that are used as primary flux calibration standards for {\it HST}.  Absorption lines of the hydrogen Balmer series are prominent in the spectrum of each star.  The fluxes are in $f_{\lambda}$ units, normalized to have a median value of 1 in the range 3850--5400~\AA\ in the top and lower-left panels. The lower-right panel shows the model fit for GD\,71 when the continuum shape is rectified. Note that the spectra and model fits are essentially indistinguishable in the plots.
From \citet*{ch1-allende09}, reproduced by permission of Wiley-Blackwell. 
\label{f:ch1-fig1}
}
\end{figure}

\begin{figure}
\figurenum{2}
\plotone{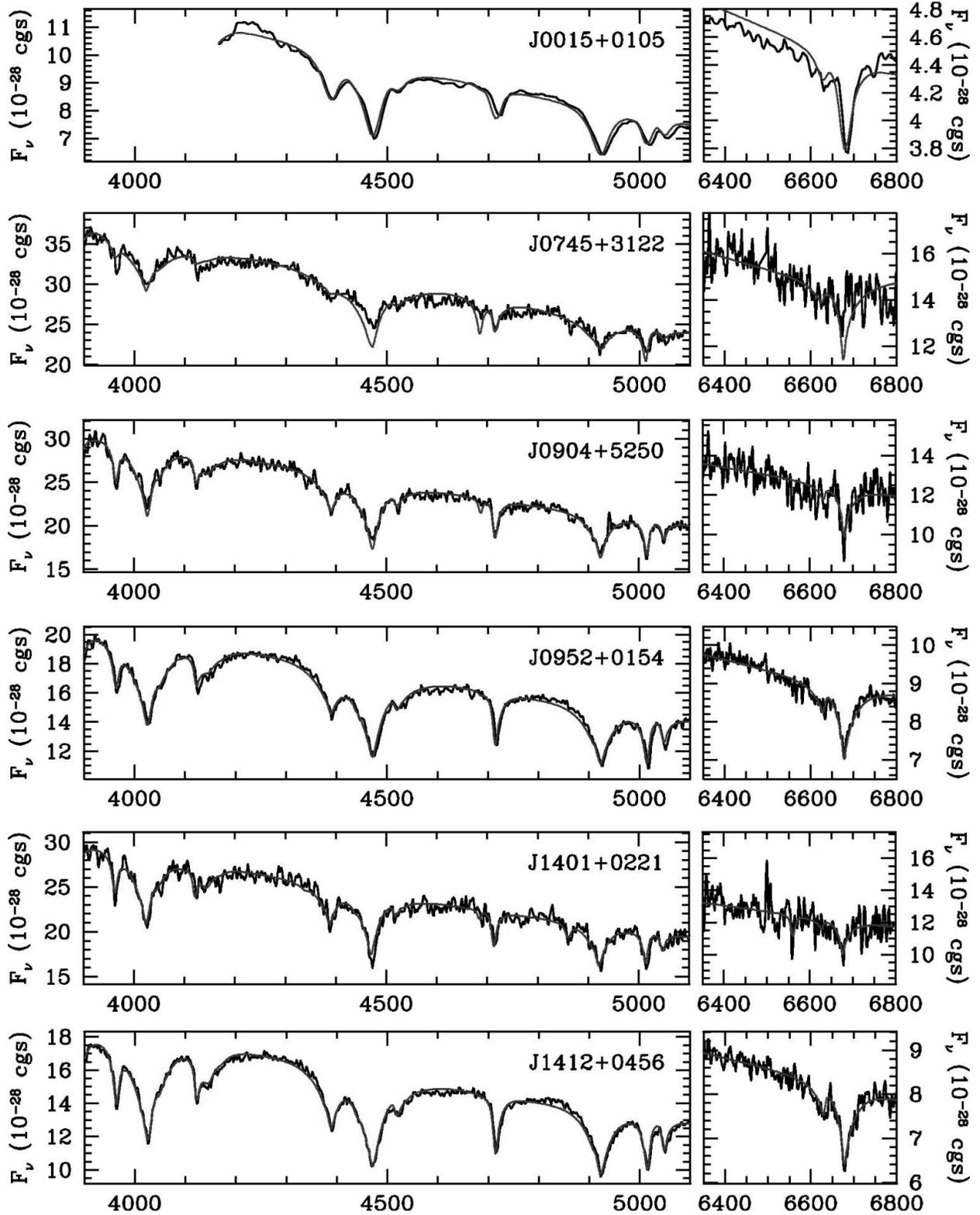}
\caption{
Spectra and model fits (smooth lines) of six hot DB white dwarfs, with temperatures in the range 30,000--45,000~K.  The prominent absorption features in each spectrum are lines of He~I.
From \citet{ch1-eisenstein06b}, reproduced by permission of the AAS.
\label{f:ch1-fig2}
}
\end{figure}

The DA stars are much easier to classify because the Balmer lines of hydrogen are detectable across a vast range of $T_{\rm eff}$, from 4000~K up to 120,000~K and higher.  Figure~\ref{f:ch1-fig1} shows representative optical spectra of several DA white dwarfs.  Hot DA stars that contain detectable helium are classified as DAO if He~II is present and DAB if He~I is present. Because of the importance of temperature as a direct luminosity and age indicator in white dwarfs, and the fact that white dwarfs span enormous ranges of $T_{\rm eff}$ (e.g., the H-rich white dwarfs span a temperature range from 4500~K to 170,000~K!), a temperature index was introduced by \citet{ch1-sion83} defined as $10\times\theta_{\rm eff} (=50,400/T_{\rm eff})$.  Thus, for the hot DA and non-DA stars, their spectral types can be expressed in half-integer steps as a function of temperature; for example, the DA sequence extends from DA.5, DA1, DA1.5, DA2, DA2.5...DA13.  A DA2 star has a temperature in the range 22,400--28,800~K, while a DA2.5 has $T_{\rm eff}$ in the range 18,327--22,400~K.  Similarly, the sequence of DB stars extends from DB2, DB2.5, DB3, DB3.5, and so on.  A DB2 star has a temperature in the range 22,400--28,800~K, while a DB2.5 has $T_{\rm eff}$ in the range 18,327--22,400~K.  Figure~\ref{f:ch1-fig2} shows representative optical spectra of several DB white dwarfs.  For the hot DA stars ($T_{\rm eff} >$ 20,000~K), the temperature index ranges are given in Table~\ref{t:ch1-tindex}.  A similar range of temperature index is defined for the hot non-DA stars.

The spectroscopic appearance of the DO and DB subclasses is determined by the ionization balance of He~I and He~II.  The DO white dwarfs show a pure He~II spectrum at the hot end, and a mixed He~I and He~II spectrum at the cool end.  However, the hottest non-DA stars are problematic to classify because (1) many are planetary nebula nuclei and isolated post-AGB stars sharing the hallmark spectroscopic characteristics of the PG\,1159  degenerates but with gravities lower than $\log{g}=7$, which is traditionally adopted as the minimum defining gravity for classification as a white dwarf star (versus a high gravity subdwarf; \citealt{ch1-gs74}); and (2) the assignment of the primary spectral class for a white dwarf is determined by the element represented by the strongest absorption features in the optical spectrum.  However, by this criterion, the PG\,1159 stars, for which either oxygen (e.g., O~VI) or carbon (e.g., C~IV) are the strongest optical lines (with He~II features weaker), should be classified as DZQO or DQZO depending upon whether O~VI or C~IV are strongest, respectively.  Since many of these objects have atmospheric compositions which are not completely helium-dominated (cf., \citealt{ch1-wh91,ch1-whh91}), it is inappropriate to assign spectral type DO on the \citet{ch1-sion83} scheme since the primary O-symbol denotes a helium-dominated composition.  Hence, the primary spectroscopic type is adopted here as the atom or ion with the strongest absorption features in the {\it optical} spectrum, where applicable (for example, it is possible that the strongest absorption feature may lie in the ultraviolet).  This is the scheme used for classifying the hottest degenerates.

In practice, a degenerate classification is withheld for any PG\,1159 star with $\log{g}<7$.  However, these objects are designated PG\,1159 as given by \citet{ch1-wh91} and \citet{ch1-whh91}, and subsequently used by \citet{ch1-nap91} and \citet*{ch1-dwh95}.  These designations are:\ E for emission, lgE for low  gravity with strong central emission, A for absorption, Ep for emission/peculiar, and EH, lgEH or AH for hybrid PG\,1159 stars that have detectable hydrogen.  The temperature index would differentiate the hot C-He-O stars from the well-known, very much cooler DZ and DQ degenerates below 10,000~K.  For example, PG\,1159 itself ($\log{g}=7$, $T_{\rm eff} =$ 110,000~K, C~IV absorption as the strongest optical lines) would be classified as DQZO.4.  The obvious disadvantage is the inevitable confusion with the cool DQ and DZ degenerates in cases for which the temperature index is missing or there are no He~II absorption features (e.g., H\,1504; \citealt{ch1-nousek86}).  In a case like H1504, where no helium is detected, $T_{\rm eff} =$ 170,000~K, and $\log{g}=7$, the classification DZQ.3 is assigned. 

The hot DQ stars are an entirely new subclass of hot white dwarfs. Unlike the previously known, and cooler, DQ white dwarfs which have helium-dominated atmospheres, the hot DQ stars have atmospheres that are dominated by carbon!  These hot DQs probably evolve from a different progenitor channel than the cool DQs. While the temperature index can be used to distinguish these C-dominated objects from the cooler He-dominated DQ stars, their uniqueness suggests a special classification designation as ``hot DQ'' stars defined by the dominant presence of C~II features in their optical spectra. There is certainly a precedent for having a special designation for this unique class of C-dominated white dwarfs.  The PG\,1159 stars merited their own special designation, rather than classifying them as DZ, DQ, or DO based upon whether O, C, or He features dominated their optical spectra. The designation PG\,1159 is widely used to distinguish these unique, exotic objects from helium-dominated DO stars at lower temperatures.  See the chapter by P.\ Dufour in this volume for details of the hot DQ white dwarfs.

\section{The Hot DA Stars}

The total number of hot ($T_{\rm eff} >$ 20,000~K) DA white dwarfs has increased enormously due in largest part to the Sloan Digital Sky Survey (SDSS; \citealt{ch1-york00,ch1-eisenstein06a}) but also to smaller surveys such as the Supernova Progenitor SurveY (SPY) project \citep{ch1-koester01,ch1-nap03}, and the Hamburg-Schmidt \citep{ch1-hagen95,ch1-homeier03} and Montreal-Cambridge-Tololo surveys \citep{ch1-demers86,ch1-lamontange97}.  Follow-up high quality ground-based spectroscopy of survey objects yield large samples of hot DA white dwarfs with precise temperatures and gravities. Only one of many such examples is the \citet{ch1-koester09} analysis of 615 DA white dwarfs from the SPY project. 

When the Balmer lines are extremely weak due to ionization, it is difficult to determine accurate temperatures for the hottest DA white dwarfs. While the Lyman series can be used to measure $T_{\rm eff}$ and estimate $\log{g}$ in the far-ultraviolet with Orfeus (e.g., \citealt{ch1-dup98}), the Hopkins Ultraviolet Telescope (HUT; e.g., \citealt{ch1-kruk97}), and {\it FUSE\/} \citep{ch1-sahnow00}, there are two different widely known discrepancies that plagued the reliable determination of hot DA physical parameters:\ (1) an inability to fit all of the Balmer lines simultaneously with consistent atmospheric parameters (the so-called Napiwotzki effect; cf.\ \citealt{ch1-gian10} and references therein); and (2) the disagreement between the parameters derived from fitting optical spectra and those derived from fitting far-ultraviolet spectra (e.g., \citealt*{ch1-finley97} and references therein). The Napiwotzki effect has been resolved by adding metals (not detected in the optical spectra) to the model atmospheres, which provides a mild back-warming effect. The fact that the analysis of far-ultraviolet spectra from the FUSE archive reveals a correlation between higher metallic abundances and instances of the Balmer line problem strongly supports this scenario \citep{ch1-gian10}.  However, the disagreement between parameters derived from optical and far-ultraviolet spectra remains.    

A large fraction of the hot DA stars observed in the extreme- and far-ultraviolet have revealed trace abundances of numerous heavy elements which are presumably radiatively levitated against downward diffusion by radiative forces \citep{ch1-chayer95}.  Extreme-ultraviolet observations of hot DA white dwarfs have been particularly effective in revealing levitated trace metals in their atmospheres \citep{ch1-finley97}. This occurs because the extreme-ultraviolet flux of a hot DA star can be strongly suppressed due to both the low opacity of the residual neutral hydrogen shortward of 300~\AA, and the strong continuum absorption and heavy line blanketing in that same extreme-ultraviolet wavelength range due to any trace metal ion constituents that may be present in the photosphere. \citet{ch1-finley97} point out an extensive literature on extreme-ultraviolet analyses of hot DA white dwarfs including, for example, work by \citet{ch1-kahn84}, \citet*{ch1-petre86}, \citet{ch1-jordan87}, \citet{ch1-PH89}, \citet{ch1-barstow93}, \citet{ch1-finley93}, \citet{ch1-jordan94}, \citet{ch1-vennes94}, \citet{ch1-vennes96}, \citet*{ch1-wolff96}, and \citet{ch1-marsh97}.

Absorption features due to C, N, O, Si, Fe, and Ni, have been seen, and in one object, PG\,1342+444, absorption lines of O~VI are detected which had previously only been seen in the subluminous Wolf-Rayet planetary nuclei and the PG\,1159 stars \citep{ch1-barstow02}.  On the other hand, there are also a sizable number of hot DA stars that appear to be metal deficient since radiative forces theory \citep{ch1-chayer95} predicts that any metals present should be levitated.  

There are now well over 100 known DA white dwarfs with $T_{\rm eff} >$ 60,000~K (e.g., \citealt{ch1-kidder91,ch1-bergeron94,ch1-finley97,ch1-marsh97,ch1-vennes97,ch1-homeier98,ch1-vennes99,ch1-koester01,ch1-liebert05,ch1-eisenstein06a,ch1-voss06,ch1-koester09,ch1-limoges10}).  Table~\ref{t:ch1-hotwds} lists a selection of these objects with the most reliable temperature estimates hotter than 70,000~K.  Noteworthy among this sample are (i) a uniquely massive hot DA white dwarf, WD\,0440$-$038, with $T_{\rm eff}$ estimates in the range 65,000~K -- 72,000~K and $\log{g}$ in the range 8.4--9.1, corresponding to a mass of $M_{\rm wd} \approx$ 0.9--1.3~$M_{\odot}$ \citep{ch1-finley97,ch1-vennes97,ch1-dup02,ch1-koester09}; and (ii) the fact that the DA stars reach temperatures above 100,000~K, with the hottest currently known DA star, WD\,0948+534, at $T_{\rm eff} \gtrsim$ 130,000~K \citep{ch1-liebert05}.  Thus, the hottest DA stars overlap in temperature with the hot DO stars and the PG\,1159 stars. This overlap further strengthens the case for the existence of a separate evolutionary channel for the hot DA stars extending up to the domain of the H-rich central stars of planetary nebulae.

The large increase in the number of hot DA stars found in the SDSS has spawned new insights into the luminosity function of hot white dwarfs. A study by \citet{ch1-krzes09} used the new SDSS sample of hotter, fainter DA stars \citep{ch1-eisenstein06a} to derive a hot white dwarf luminosity function that extends to the most luminous, hottest white dwarfs to date.  This luminosity function encompasses DA and non-DA stars over the temperature range $\sim$25,000~K $> T_{\rm eff} > \sim$ 120,000~K.  The cool end of their luminosity function connects with the hot end of previously determined SDSS white dwarf luminosity functions.  By constructing separate DA and non-DA luminosity functions for the first time, \citet{ch1-krzes09} noted distinct differences between them.  For example, they found a sudden drop in the non-DA luminosity function near $M_{\rm bol} = 2$, which they interpret as a transition of non-DA atmospheres into DA atmospheres during white dwarf evolution. The transition would occur as trace amounts of hydrogen float to the surface and give rise to H-features in the optical spectrum.

It is well known that the hottest DA stars are cooler than the hottest non-DA stars \citep{ch1-sion86,ch1-fontaine87}.  Unless some fraction of the PG\,1159 degenerates undergo spectral evolution into hot DA stars when previously ``hidden'' hydrogen floats up to their surface, then the progenitors of the hottest DA stars represent a separate evolutionary channel. This would be contrary to the single channel scenario of \citet{ch1-fontaine87} In the Hertzsprung-Russell (H-R) diagram, the hottest DA white dwarfs appear to connect up with the H-rich central stars of planetary nebulae. There is very likely a direct evolutionary connection.

\subsection{DAO White Dwarfs} 

The DAO stars are hot DA stars that exhibit ionized helium in their optical spectra.  They are characterized by surface temperatures in the range 50,000~K $< T_{\rm eff} <$ 100,000~K, low gravities with $\log{g} < 7.5$, and $\log{[{\rm He}/{\rm H}]} = -2$ by number \citep{ch1-bergeron94,ch1-nap99,ch1-good04,ch1-good05}.  Searches for binarity as a means of explaining the presence of He~II in a DA star as a composite spectrum have been largely negative \citep{ch1-good05}. There are six DAOs in binaries including four out of 12 low mass DAOs that are in binaries. Thus, the majority of the low mass DAOs appear to {\em not} be in binaries and, hence, may be the descendants of extended horizontal branch progenitors.  Because pure DA and DB stars have monoelemental atmospheres, the hybrid composition DAO stars offer insights into white dwarf spectral evolution. 

There is typically a large discrepancy in the $T_{\rm eff}$ values  determined for DAO white dwarfs from optical Balmer line spectra compared to temperatures derived from Lyman line spectra obtained with {\it FUSE\/} (see Table~\ref{t:ch1-hotdao}).  This behavior echoes the Balmer line versus Lyman line temperature discrepancy noted by \citet{ch1-barstow03} for the hot ($T_{\rm eff} >$ 50,000~K) DA white dwarfs.  The spectra of DAO stars are generally better fitted with models utilizing homogeneous, rather than stratified, atmospheres, which led \citet{ch1-good04} to suggest the possibility that homogeneous atmospheres are a better approximation of reality in low gravity DAO white dwarfs.

Virtually all of the DAO stars reveal heavy element absorption lines in their spectra. The metal abundances in the DAO stars appear to be generally higher than the metal abundances in their DA counterparts, although \citet{ch1-good04} noted little difference between the metal abundances of DAO stars and DA stars. Moreover, \citet{ch1-good05} found that the metal abundances of the lower gravity, lower temperature DAO stars and the higher gravity DAO stars differed little from each other despite the fact that these two groups of DAO white dwarfs presumably have arisen from different progenitor channels (see below).  Good et al.\ also noted that the metal abundances of the DAO white dwarfs were not markedly different from those of the hot DA stars with metals.  The observed DAO metal abundances were compared with the theoretical predictions of \citet{ch1-chayer95}, as well as with the wind mass loss calculations of \citet{ch1-unglaub98,ch1-unglaub00}.  Except for Si, none of the metal abundances exceeded the theoretical prediction after taking into account radiative levitation and weak mass loss.

The low masses of DAO white dwarfs would point toward evolution from stars that did not have sufficient mass to ignite core helium burning on the horizontal branch. Hence, the lower mass DAO stars cannot be the descendants of post-AGB evolution. It is possible that they are the descendants of the sdB and sdO subdwarf stars on the extended horizontal branch.  On the other hand, the more massive DAO stars with $M_{\rm wd} > 0.5$~M$_{\odot}$ appear to represent an evolutionary channel connecting them to the AGB stars, since DAO white dwarfs more massive than $0.5$~M$_{\odot}$ would have been massive enough for helium shell burning on the horizontal branch. In this scenario, the DAO stars could also be the progeny of H-rich planetary nebula central stars or even hybrid PG\,1159 stars (containing some H) in which a ``hidden'' reservoir of hydrogen floats up to the surface.  Recently, \citet{ch1-gian10} contend that the post-extreme horizontal branch evolution is no longer needed to explain the evolution for the majority of the DAO stars, and that the presence of metals might drive a weak stellar wind, which, in turn, could explain the presence of helium in DAO white dwarfs.  Nevertheless, it is still not possible to definitively establish these different potential evolutionary links.

\section{The PG\,1159 Stars}
\label{s:ch1-pg1159}

The most exciting stellar discovery of the Palomar Green survey was a class of extremely hot, high luminosity degenerate objects known as the 1159 stars \citep*{ch1-green86}.  Subsequently, large surveys (Palomar Green; Hamburg-Schmidt; Hamburg ESO, \citealt{ch1-wis96}; and, most recently, the SDSS) have uncovered the majority of the known PG\,1159 stars\footnote{The most recent and only discovery within the last 10 years, besides those from the SDSS, is HE\,1429$-$1209, which was discovered as a white dwarf candidate in the Hamburg ESO survey and confirmed as a PG\,1159 star by the SPY project \citep{ch1-werner04}.}.  The PG\,1159 stars reveal spectra typically devoid of hydrogen.  Instead, they are dominated by He~II and highly ionized, high excitation carbon, especially a broad absorption trough in the region of 4670~\AA\ comprising He~II $\lambda4686$ and several high excitation C~IV lines \citep{ch1-green86}.  Absorption lines of O~VI were also detected in the optical ultraviolet \citep*{ch1-sion85}.  All of these high excitation features implied very high effective temperatures. However, the first determinations of accurate surface temperatures and the first chemical abundances for these extremely hot objects had to await the development of a new generation of Non-Local Thermodynamic Equilibrium (NLTE) atmosphere codes with the requisite atomic data for millions of transitions (by Klaus Werner and his students; discussed below).  Meanwhile, the SDSS made it possible to significantly increase the number of known PG\,1159 stars.

Synthetic spectral analyses require studies of lines of successive ionization stages in order to evaluate the ionization equilibrium of a given chemical element. This provides a sensitive indicator of the effective temperature. Since stars with $T_{\rm eff}$ as high as 100,000~K have their Planckian peaks in the extreme ultraviolet wavelength range and the features are highly ionized, most of the metal lines are found in the ultraviolet range. Werner and his colleagues require high signal-to-noise and high resolution ultraviolet spectra for their NLTE analyses of white dwarfs.  They utilized a number of spacecraft instruments including the Faint Object Spectrograph, Goddard High Resolution Spectrograph, Space Telescope Imaging Spectrograph, and Cosmic Object Spectrograph onboard {\it HST\/} in order to acquire spectra of sufficient quality.  They carried out state-of-the-art analyses of the hottest (pre-)white dwarfs by means of NLTE model atmospheres, which include the metal-line blanketing of all elements from hydrogen to nickel \citep{ch1-rauch10}.

\begin{figure}
\figurenum{3}
\plotone{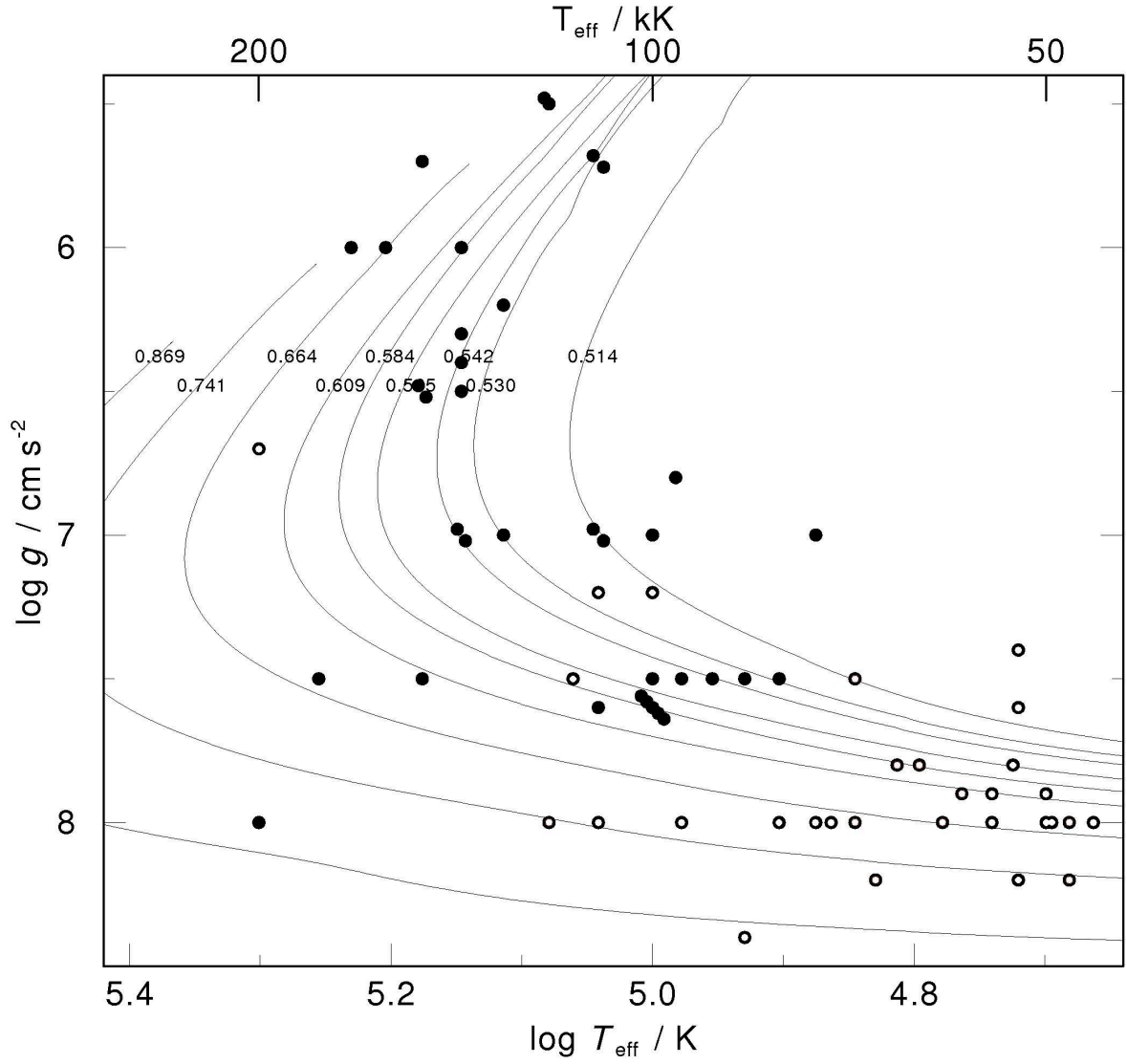}
\caption{
Summary plot of $\log{g}$ versus $\log{T_{\rm eff}}$ showing all analysed PG\,1159 stars (filled circles) and hot DO white dwarfs (unfilled circles), with DO white dwarf cooling tracks by \citet{ch1-althaus09}.
\label{f:ch1-fig3}
}
\end{figure}

The spectral analysis of the PG\,1159 stars revealed a range of temperatures of $T_{\rm eff} =$ 75,000--200,000~K and gravities of $\log{g} =$ 5.5--8.0 \citep{ch1-rauch91,ch1-dreizler94,ch1-werner96}.  Prior to the SDSS, only 28 PG\,1159 stars were known.  From the empirically derived ranges of their parameters, it was obvious that the position of some of the PG\,1159 stars (specifically the lower gravity members, subtype lgE; \citealt{ch1-werner92}) in the H-R diagram overlapped the hot central stars of planetary nebulae.  The remainder of the PG\,1159 stars are more compact objects with higher (white dwarf) surface gravities (subtype A or E).  Thus, the PG\,1159 stars appear to be evolutionary transition objects between the hottest post-AGB and white dwarf phases.  Figure \ref{f:ch1-fig3} shows a summary plot of $\log{g}$ versus $\log{T_{\rm eff}}$ from the most recent analyses of the PG\,1159 and hot DO white dwarfs.  For comparison, evolutionary tracks by \citet{ch1-althaus09} are also plotted.

\begin{figure}
\figurenum{4}
\plotone{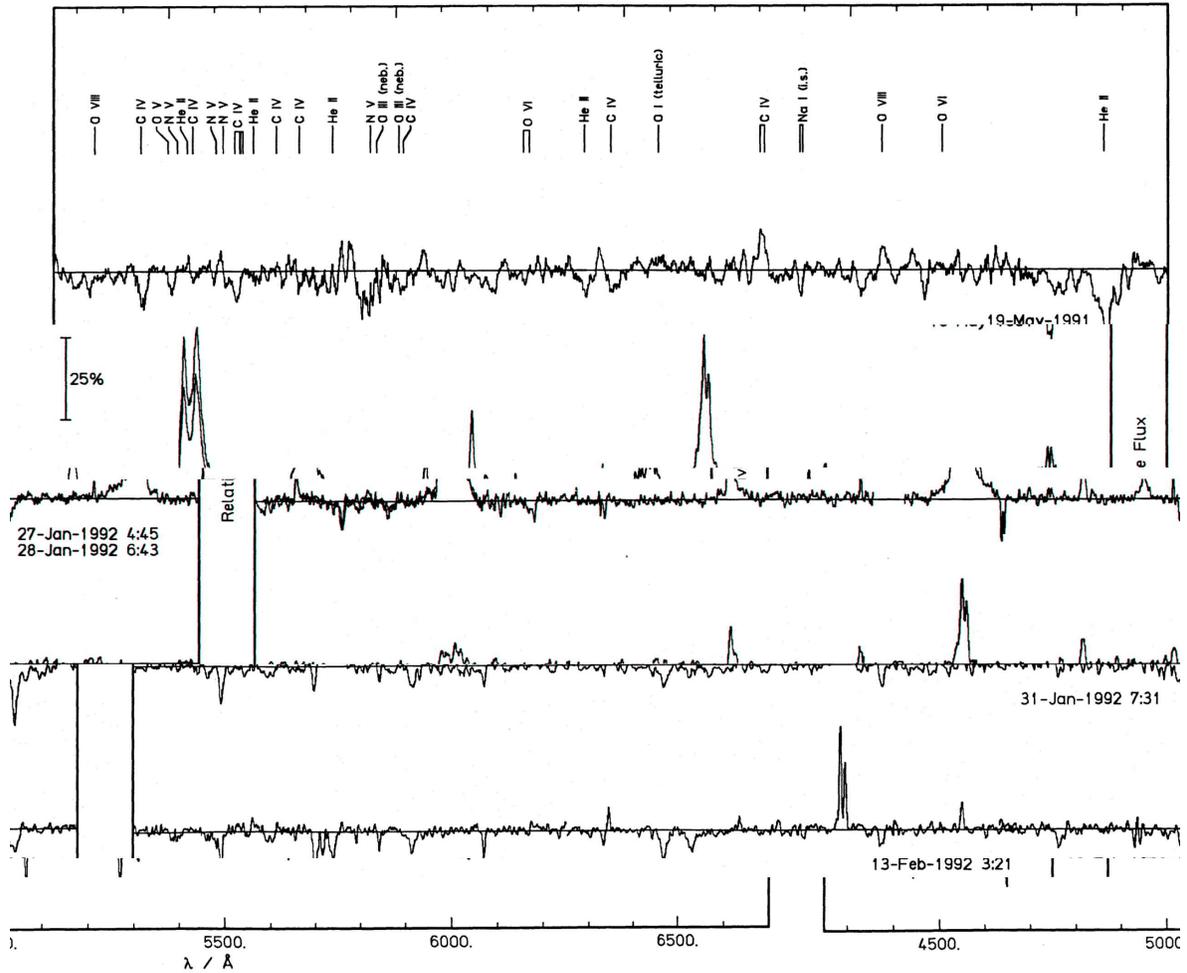}
\caption{
Time sequence of normalized spectra of the central star in Longmore~4, showing (from top to bottom) the appearance and rapid decline of a dramatic emission line phase linked to a mass ejection event. 
From \citet{ch1-longmore4}, reproduced with permission \copyright~ESO. 
\label{f:ch1-fig4}
}
\end{figure}

Until very recently, Fe lines had never been detected in PG\,1159 stars even though Fe lines are seen in some hot DA, DO, and DAO white dwarfs.  In the far-ultraviolet, features due to Fe~VII would be the expected indicators for the presence of Fe, but only if the effective temperature is not so high that the population of Fe~VII is too much depleted by ionization.  Using the plethora of far-ultraviolet transitions lying in the {\it FUSE\/} range, \citet*{ch1-werner10} have detected Fe features not only in the cooler PG\,1159 stars with $T_{\rm eff} <$ 150,000~K, but also among objects in the hottest subset of PG\,1159 stars with temperatures between 150,000~K and 200,000~K, including RX\,J2117.1$+$3412, K\,1–16, NGC\,246, H1504$+$65, and Longmore~4 (which has revealed evidence of episodic mass ejection, as shown in Figure~\ref{f:ch1-fig4}; \citealt{ch1-longmore4}).  Among the cooler subset of PG\,1159 stars, Fe~VIII lines are detected at solar abundance in {\it FUSE\/} spectra, while in the hotter subset, Fe~X is the detected species and the analysis of abundances are in progress \citep{ch1-werner10a,ch1-longmore4}.  Their analyses yielded a solar iron abundance for these stars.  These hottest objects are among the most massive PG\,1159 stars (0.71--0.82 $M_{\odot}$), while those objects revealing the strongest Fe deficiency are associated with a lower mass range (0.53--0.56 $M_{\odot}$).  Nonetheless, the evolutionary significance, if any, of the presence of Fe in solar abundance in some PG\,1159 stars versus PG\,1159 stars that appear to be Fe-deficient, remains unclear.  

It is now widely believed that the hydrogen-deficiency in extremely hot post-AGB stars of spectral class PG\,1159 is probably caused by a (very) late helium-shell flash or an AGB final thermal pulse \citep{ch1-iben84} that consumes the hydrogen envelope, exposing the usually-hidden intershell region. Thus, the photospheric elemental abundances of these white dwarfs offer insights into the details of nuclear burning and mixing processes in the precursor AGB stars.  \citet{ch1-werner08}  compared predicted elemental abundances to those determined by quantitative spectral analyses performed with advanced NLTE model atmospheres. A good qualitative and quantitative agreement is found for many elemental species (He, C, N, O, Ne, F, Si, Ar), but discrepancies for others (P, S, Fe) point at shortcomings in stellar evolution models for AGB stars.  PG\,1159 stars appear to be the direct progeny of [WC] Wolf-Rayet stars \citep*{ch1-werner07,ch1-werner08}, a possibility first suggested by \citet{ch1-sion85} when the same high excitation O~VI absorption features detected in the PG\,1159 stars were also seen in the optical ultraviolet spectra of O~VI planetary nebula nuclei.

\section{DO White Dwarfs}

\begin{figure}
\figurenum{5}
\plotone{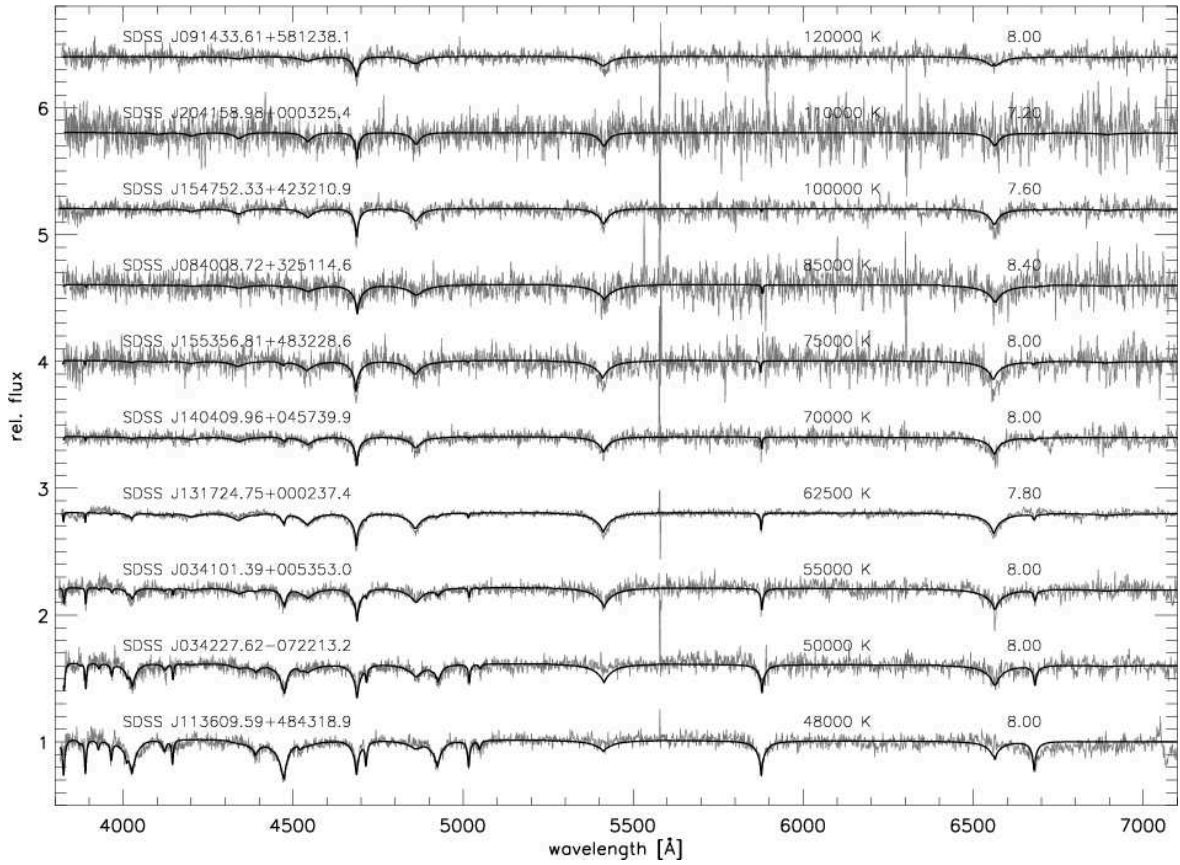}
\caption{
Normalized optical spectra (grey lines) of DO white dwarfs discovered in the SDSS, along with model atmospheres (black lines), ordered by decreasing effective temperature. 
From \citet{ch1-hug05}, reproduced with permission \copyright~ESO. 
\label{f:ch1-fig5}
}
\end{figure}

The DO white dwarfs are hot helium-rich white dwarfs that populate the white dwarf cooling sequence from the hot beginning ($T_{\rm eff} =$ 120,000~K) down to 45,000~K.  The optical spectra of hot DO stars covering this range of $T_{\rm eff}$, and newly discovered in the SDSS, are displayed in Figure~\ref{f:ch1-fig5}.

At $T_{\rm eff}<$~45,000~K, the helium-rich cooling sequence is interrupted by the so-called ``DB gap'' \citep{ch1-liebert86} in which, prior to the SDSS, no helium-rich white dwarfs were found down to effective temperatures of 30,000~K.  \citet{ch1-dw96,ch1-dw97} determined effective temperatures and surface gravities of all 18 known DO white dwarfs.  The DO stars have a nearly pure helium atmosphere, with only relatively few showing weak metal lines. Their surface composition is controlled by the competition between gravitational diffusion and radiative levitation, as well as possible weak mass loss. The initial comparisons between theoretical predictions for white dwarfs and the observed abundances were obtained by \citet{ch1-fontaine79} and \citet{ch1-vvg79}, who took into account radiative levitation.  These processes are also invoked to explain the transition from the possible progenitors, the hydrogen deficient, C- and O-rich PG\,1159 stars (see \citealt{ch1-dh98}, as well as Section \ref{s:ch1-pg1159} above, for an introduction to these objects).  In this scenario, heavier elements in the atmosphere are depleted and only traces of metals can be kept in the helium rich atmosphere as long as the radiative levitation is sufficiently strong due to high effective temperatures.  Finally, at the hot end of the DB gap, enough of the remaining traces of hydrogen floats up to cover the helium rich envelope with a thin hydrogen rich layer. Even though this scenario is able to qualitatively explain the evolution of hot helium rich white dwarfs, these processes are far from understood quantitatively.  For example, observed metal abundances (see Table~\ref{t:ch1-do}) do not fit theoretical predictions \citep{ch1-chayer95,ch1-dw96,ch1-dw97}.  Whether this is due to the rather poor observational data, lack of adequate models, or a fundamental problem with the transition scenario itself has so far not been determined.

From the analysis of PG\,1159 stars it is known that the nitrogen abundance varies at least by three orders of magnitude \citep{ch1-dh98}.  Are the PG\,1159 stars the progenitors of the DO white dwarfs?  Since nitrogen is destroyed in triple alpha reactions, it is not surprising that the PG\,1159 stars are generally nitrogen-poor.  Hence, it is possible that those DO stars with low nitrogen abundance could be the direct descendants of PG\,1159 stars with low nitrogen abundance.

Among the DO white dwarfs are exotic objects with ultra-high excitation ion (uhei) lines in their optical spectra.  These objects were first discovered prior to the SDSS. They show absorption lines of O~VII and N~VII around 3888~\AA, at 5673~\AA, and around 6086~\AA, as well as O~VIII at 4340~\AA, 4658~\AA, and 6064/6068~\AA.  In a new SDSS uhei DO white dwarf, SDSS~J025403.75$+$005854.4, features due to Ne~IX are also present \citep{ch1-krzes04}.  The He~II lines in these uhei DO white dwarfs are very strong and cannot be fitted with the latest NLTE models. The effective temperature required to excite the detected ions exceeds 500,000~K.  The evolutionary track of a massive (1.2 $M_{\odot}$) post-AGB remnant carries it to effective temperatures as high as 700,000~K on very short time scales \citep{ch1-pac70}. However, \citet{ch1-werner95} have argued that the uhei features in DO white dwarfs cannot be photospheric because the He~II lines would fade completely at such high temperatures.  Instead, they proposed an hypothesis of optically thick and hot stellar winds, based on the triangular shape of the line profile. Furthermore, the lines are blue-shifted, which favors the assumption of an expanding envelope. 

Related to these uhei DO stars is KPD\,0005$+$5106, one of the hottest know DO stars.  \citet{ch1-werner07,ch1-werner08} discovered highly ionized photospheric metals, such as Ne~VIII and Ca~X, requiring extremely high temperature, much higher even than previous analyses that yielded $T_{\rm eff}\sim$ 120,000~K.  More recently, \citep{ch1-wasserman11} reported the detection of Si~VII, S~VII, and Fe~X in this star.  The NLTE analysis of the metals and the He~II line profiles yield $T_{\rm eff}=$ 200,000~K with $\log{g}=6.7$.  The abundances of metals are in the range of 0.7 to 4.3 times solar with an upper limit to any hydrogen present of $<0.034$ solar.  Remarkably, these new analyses of KPD\,0005$+$5106 reveal it to be very much hotter than the next hottest DO white dwarf. In the H-R diagram, it stands alone, far to the left and at higher luminosity than any of the other DO stars.  At its high $T_{\rm eff}$ and luminosity it is likely that the chemical abundances are probably affected by a stellar wind.  Thus, diffusion and radiative levitation may not be important factors in controlling the surface abundances. Furthermore, \citet{ch1-wasserman11} found that the chemical abundances of KPD\,0005$+$5106 most closely resemble the abundances seen in R~Coronae~Borealis stars.  Since the R~Cor~Bor objects are widely held to be the product of binary mergers (e.g., \citealt{ch1-webbink84}, \citealt{ch1-han98}), this may imply that KPB\,0005$+$5106 is itself the product of such a merger and, hence, is the evolved progeny of an R~Cor~Bor giant.  If true, such a connection would imply that the surface abundances of KPD\,0005$+$5106 are chemical relics of the progenitor giant and, thus, not controlled by diffusion and radiative levitation.  If this interpretation is correct, then KPD\,0005$+$5106 would represent a new evolutionary channel producing DO white dwarfs distinct from the evolutionary channel connecting the PG\,1159 stars to the DO stars.

\section{DB White Dwarfs}

The DB stars contain helium to a degree of purity not seen in any other astronomical objects. Even at high signal-to-noise ratio, many spectra exhibit only the absorption lines of He~I.  If the DB star has accreted metals from a debris disk, comets, or the interstellar medium, then they are classified DBZ, or DBAZ if hydrogen is present (see below). Some hot DB stars exhibit atomic carbon in the far ultraviolet. The best example is the pulsating hot DB star GD\,358 \citep{ch1-sion89}.  It remains unclear if the DBA stars have accreted their hydrogen or if it is primordial and a result of convective mixing.  The DB white dwarfs by consensus are the progeny of extremely hydrogen deficient post-AGB stars (e.g., see \citealt{ch1-althaus05,ch1-althaus09}, and references therein).  The DB cooling sequence extends from the hottest DB stars like GD\,358 and PG\,0112$+$122, down to the cooler DB white dwarfs (below 20,000~K), and extending downward to 12,000~K, at which point envelope convection has deepened substantially and the He~I lines become undetectable.  The distribution by number of the coolest DO stars to the hottest DB stars (30,000~K to 45,000~K) is interrupted by the DB gap \citep{ch1-liebert86}.  Prior to the SDSS, within this very wide range of temperature, no objects with H-deficient atmospheres were known to exist. 

Now, however, the large number of new white dwarfs discovered in the SDSS \citep{ch1-eisenstein06a} has led to the firm placement of no less than 26 DB stars within the DB gap \citep{ch1-hug09}.  This raises suspicions that the DB gap was not a real feature of the white dwarf temperature distribution.  On the other hand, there is still a deficit of a factor of 2.5 in the DA/non-DA ratio within the gap \citep{ch1-eisenstein06b}.  However, many other objects whose status is questionable (e.g., DAO, DBA, DAB, masquerading composite DA+DB/DO systems) may alter or eventually erase this deficit.  Some of these objects are found within the gap, while others are seen near the gap edges.  There is also the complicating factor of circumstellar accretion of hydrogen, and the role played by radiative levitation and weak winds in this temperature interval. 

If every DA white dwarf evolving through the DB gap turned into a DB, then there should be a significant spike seen in the number of DB stars at the red edge of the DB gap, which is not observed. This is in contrast to the strong signature of convective mixing and dilution that changes (significantly lowers) the DA to non-DA ratio at lower temperatures, $T_{\rm eff} <$ 12,000~K \citep{ch1-sion84}.  Even if the transformation of DA stars into DB stars takes place at 30,000~K and lower, when a sufficiently thin hydrogen layer is convectively mixed downward and diluted by the deepening helium convection zone.  This only affects the DA stars with extremely low hydrogen layer masses ($<10^{-15}$~$M_{\rm wd}$), which amounts to approximately 10\% of all DA stars cooling through the 45,000~K to 30,000~K interval \citep{ch1-eisenstein06b}.  This is contrary to the original contention of \citet{ch1-fontaine87} that all DA stars should have ultra-thin layers ($<10^{-12}$~$M_{\rm wd}$).  Rather, it appears that the fraction of hot DA white dwarfs that transform into non-DA white dwarfs is on the order of 10\% of all DA stars.  

The problem of whether a DB gap exists is complicated by the known existence of several peculiar DAB, DBA, or DAO stars believed to lie in the 30,000--45,000~K range that (1) show evidence of spectrum variability and/or (2) do not fit atmosphere models, whether homogeneous (completely mixed) in H and He throughout the atmosphere or stratified with the hydrogen all in a very thin upper layer.  Several other white dwarfs in or near the DB gap also have peculiar spectra.  For example, PG\,1210$+$533, with $T_{\rm eff} =$ 45,000~K, exhibits line variability of H, He~I, and He~II, probably modulated by rotation \citep{ch1-bergeron94}.  Also, GD\,323, with $T_{\rm eff} =$ 30,000~K, is a DAB star with a variable spectrum that cannot be fit completely successfully by either homogeneous or stratified model atmospheres \citep*{ch1-liebert84, ch1-pereira05, ch1-koester09}.  Additional examples include HS\,0209$+$0832, with $T_{\rm eff}=$ 36,000~K and a 2\% helium abundance \citep{ch1-jordon93}, and PG\,1603$+$432, with $T_{\rm eff}=$ 35,000~K and a 1\% helium abundance \citep*{ch1-vennes04}.  The existence of these systems adds to the mounting evidence that the DB gap is not real, as increasing numbers of He-rich stars are being discovered within and near its boundaries.

\subsection{DBA White Dwarfs}

The DBA stars have strong lines of He~I and weaker Balmer lines, hydrogen-to-helium ratios (by number) in the range $N({\rm H})/N({\rm He})\sim10^{-5}$ to $10^{-3}$, and, for the most part, effective temperature below 20,000~K \citep*{ch1-shipman87}.  Hence, they cluster at the low end of the DB temperature distribution.  The DBA white dwarfs were previously thought to comprise roughly 20\% of He-rich white dwarfs between 12,500~K and 20,000~K \citep{ch1-shipman87}.  However, more recent surveys have dramatically changed this picture.  The SPY project yielded a sample of 71 helium-rich degenerates, of which six were new DBA discoveries and 14 were DB stars reclassified as DBA due to the detection of hydrogen lines \citep{ch1-voss07}.  In all, 55\% of their SPY sample were DBA stars.  This is a factor of almost 3 times higher than the fraction of DBA stars first estimated by \citet{ch1-shipman87}.

It remains unclear if the DBA stars have accreted their hydrogen or if the small H mass was originally primordial, diluted by convection, and then floated back to the surface as a result of convective mixing.  This large fraction of DBA stars, coupled with the total hydrogen masses estimated for the DBA stars suggests the possibility that DB white dwarfs, as they cool, accrete interstellar hydrogen, thus raising their hydrogen mass to the point at which a DBA star appears.  This channel for forming DBA white dwarfs was favored if the DB gap was real, because DB white dwarfs could be masquerading as DA stars in the DB gap with only a thin hydrogen layer ($<10^{-15}$ $M_{\rm wd}$).  This layer could be mixed away and diluted, resulting in a DB star appearing at the cool end of the DB gap. It now appears that there is no DB gap.  Hence, this constraint on the H-layer mass is no longer relevant.  While interstellar accretion of hydrogen cannot be easily dismissed, it may yet prove plausible that the hydrogen is accreted from volatile-rich debris or comets. The fact that there are DBAZ and DBZA stars with accreted calcium may support this scenario.

\section{Hot DQ White Dwarfs}

The discovery of hot DQ white dwarfs with carbon-dominated atmospheres (\citealt{ch1-dufour07,ch1-dufour08}; also see the chapter in this volume by P.\ Dufour).  in the SDSS Data Release 4 sample has raised new questions about white dwarf formation channels. These objects are distinctly different from the cooler, normal DQ stars, which have helium-dominated atmospheres and carbon abundances of $\log{N({\rm C})/N({\rm He})}\sim10^{-5}$ by number, with the highest carbon abundance measured for ordinary DQ stars being $\log{N({\rm C})/N({\rm He})}\sim10^{-3}$.  The hot DQs all have effective temperatures between $\sim$18,000~K and 24,000~K.  Their surface compositions are completely dominated by carbon, with no evidence of H or He~I in their optical spectra. Their surface gravities are typically $\log{g}=8$, with one object (SDSS\,J142625.70$+$575218.4) having a gravity near $\log{g}=9$ \citep{ch1-dufour08}.  Their optical spectra contain numerous absorption lines of C~II, which is the hallmark spectroscopic signature of the hot DQs. Among the strongest transitions of C~II are at 4267~\AA, 4300~\AA, 4370~\AA, 4860~\AA, 6578~\AA, and 6583~\AA. 

Despite extensive searches of the vast SDSS sample, no carbon-dominated DQ stars have been found with $T_{\rm eff}$ higher than the hottest hot DQ white dwarf, at 24,000~K.  Based upon this absence of hotter carbon-dominated DQ stars, it is quite possible that these objects appear helium-dominated at higher temperatures, but with very low mass helium layers that could be effectively mixed and diluted in the carbon-rich convection zone that forms and deepens due to carbon recombination as the hot DQ star cools \citep{ch1-dufour08}.  However, the helium layer would have to be thin in order to be hidden from spectroscopic detection as the hot DB transforms into a hot DQ star. Adding to the puzzle posed by the hot DQ stars, \citet{ch1-dufour09} point out an exceptionally high fraction of hot DQ stars with high magnetic fields ($\sim$40--60\% among the hot DQ white dwarfs, compared with $\sim$10\% for the sample of nearby white dwarfs of all types; \citealt{ch1-liebert03}). 

It seems likely that these objects could be cooled-down descendants of stripped carbon core objects like H\,1504$+$65 \citep{ch1-nousek86}.  They could have experienced a late thermal pulse that eliminated most of the helium, a phenomenon similar to the one that is generally believed to explain the existence of other hydrogen deficient stars \citep{ch1-wh06}.  If the hot DQ white dwarfs are cooled versions of stripped carbon-oxygen cores like H\,1504$+$001, then this would be an entirely new evolutionary channel.

\section{Conclusions}

It is clear that most of the progress achieved in understanding the competing physical processes in hot white dwarf envelopes, the spectral evolution of hot white dwarfs, and the identification of the white dwarf progenitor channels has arisen directly from an interactive combination of synthetic spectral abundances studies via space- and ground-based spectroscopy, with studies of nuclear astrophysics and thermal instabilities via AGB and post-AGB stellar evolutionary sequences including mass loss.  A major triumph has been the successful prediction of surface abundances in hot white dwarfs from born-again thermal pulse models and AGB thermal pulse models. This has led to agreement between the observed surface abundances from synthetic spectral analyses of high gravity post-AGB stars (PG\,1159 and subluminous Wolf-Rayet planetary nebula nuclei) and the theoretical intershell abundances of their double shell-burning AGB progenitors \citep{ch1-wh06}.  That pre-white dwarf, post-AGB surface abundances shed light on nuclear astrophysical processes deep inside the intershell layers of the progenitor AGB star is all the more remarkable.

While episodic mass ejection and winds are directly observed in post-AGB stars, including planetary nebula nuclei, evolving on the plateau and knee portions of pre-white dwarf evolutionary tracks, it remains unknown whether theoretically-predicted weak wind mass loss and ion-selective winds exist along the white dwarf cooling tracks \citep{ch1-unglaub08}.  Yet, along the white dwarf cooling tracks at $T_{\rm eff} <$~50,000~K, where the most marked disagreement between observed and predicted surface abundances occur, weak winds may not exist. Complicating this scenario is the possible interplay of accretion with diffusion and radiative levitation at $T_{\rm eff} >$~20,000~K. At $T_{\rm eff} >$~50,000~K, the decreasing abundances of helium and metals are consistent with the combined effects of diffusion, radiative forces, and weak wind outflow. This wind outflow has not yet been detected and associated mass loss rates are unknown \citep{ch1-unglaub08}.

The future of research on hot white dwarfs will undoubtedly involve ever more sophisticated, multi-dimensional, quasi-static and hydrodynamic evolutionary calculations with more realistic mass loss prescriptions, and a greatly expanded database of atomic data and model atoms, as well as the continued incorporation of diffusion and other envelope physical process into NLTE model atmosphere codes.  These inevitable developments, along with the anticipated quantum leap in the number of white dwarfs with measured accurate parallaxes obtained by the GAIA Mission (e.g., \citealt{ch1-gaia09}), will lead to an ever more detailed knowledge of white dwarf formation channels and the envelope physical processes governing and controlling spectral evolution down the white dwarf cooling sequence. 

An added windfall is the availability of ultra-sophisticated model atmosphere codes (e.g., TLUSTY/SYNSPEC\footnote{http://nova.astro.umd.edu}, \citealt{ch1-tlusty1}, \citealt{ch1-tlusty2}; German Astrophysical Virtual Observatory\footnote{http://www.g-vo.org/}) to a widening circle of investigators, which should also serve to enhance the quantitative analyses shedding light on the formation and spectral evolution of hot degenerate stars.

\acknowledgements
It is a pleasure to thank Jay Holberg for useful discussions of hot DA stars and Patrick Dufour for discussions of hot DQ stars.  I would also like to thank Klaus Werner for providing temperature data on hot non-DA stars in advance of publication. This work was supported by NSF grant AST1008845, and in part by NSF grant AST807892, both to Villanova University.

\clearpage

\begin{deluxetable}{l l}
\tablecolumns{2} 
\tablewidth{0pt}
\tablecaption{Definition of Primary Spectroscopic Classification Symbols \label{t:ch1-symbols}}
\tablehead{
\colhead{Spectral Type} & 
\multicolumn{1}{l}{Characteristics}
}
\startdata 
DA .............. & Only Balmer lines; no He~I or metals present \\ 
DB .............. & He~I lines; no H or metals present \\ 
DC .............. & Continuous spectrum, no lines deeper than 5\% \\
                  & ~~~ in any part of the electromagnetic spectrum \\ 
DO .............. & He~II strong; He~I or H present	\\ 
DZ .............. & Metal lines only; no H or He lines \\ 
DQ .............. & Carbon features, either atomic or molecular \\
                  & ~~~ in any part of the electromagnetic spectrum \\ \\
~~~~ P ........... & magnetic white dwarfs with detectable polarization \\ 
~~~~ H ........... & magnetic white dwarfs without detectable polarization \\ 
~~~~ X ........... & peculiar or unclassifiable spectrum \\ 
~~~~ E ........... & emission lines are present \\ 
~~~~ ? ........... & uncertain assigned classification; a colon (:) may also be used \\
~~~~ V ........... & optional symbol to denote variability
\enddata
\end{deluxetable}

\begin{deluxetable}{lcc}
\tablecolumns{3} 
\tablewidth{0pt}
\tablecaption{Temperature Index Ranges for Hot DA Stars	\label{t:ch1-tindex}}
\tablehead{
\colhead{Spectral Type} & 
\colhead{$T_{\rm eff}$ Range (K)} & 
\colhead{$10\times\theta_{\rm eff}$ Range} 
}
\startdata 
DA.25  & 200,000 & \nodata \\ 
DA.5   & 100,800 & \nodata \\ 
DA1    & 40,320--67,200 & 1.25--0.75 \\
DA1.5  & 28,800--40,320 & 1.75--1.25 \\
DA2    & 22,400--28,800 & 2.25--1.75 \\
DA2.5  & 18,327--22,400 & 2.75--2.25
\enddata
\end{deluxetable}

\begin{deluxetable}{l l l l}
\tabletypesize{\footnotesize} 
\tablecolumns{4} 
\tablewidth{0pt}
\tablecaption{A Selection of the Hottest DA White Dwarfs ($T_{\rm eff} >$ 70,000~K)\tablenotemark{a} \label{t:ch1-hotwds}}
\tablehead{
\colhead{WD Name} & 
\colhead{$T_{\rm eff}$ (K)} & 
\colhead{$\log{g}$} & 
\colhead{Reference} 
}
\startdata 
0556$-$375$^{\ast}$ &  70,275$^{+8505}_{-3875}$ & $7.37^{+0.13}_{-0.24}$ & \citet{ch1-marsh97} \\  
1248$-$278          &  70,798$\pm1305$ & $7.23\pm0.06$ & \citet{ch1-kidder91,ch1-koester01} \\
1312$-$253$^{\ast}$ &  71,153$\pm906$ & $7.06\pm0.04$ & \citet{ch1-kidder91,ch1-koester01} \\
1312$-$253$^{\ast}$ &  71,234$\pm891$ & $7.02\pm0.03$ & \citet{ch1-kidder91,ch1-koester01} \\ 
0440$-$038$^{\ast}$ &  71,545$\pm805$ & $8.538$ & \citet{ch1-voss06} \\  
2244$+$031          &  72,000$\pm1600$ & $7.78\pm0.08$ & \citet{ch1-homeier98} \\
0440$-$038$^{\ast}$ &  72,340$\pm2319$ & $8.772\pm0.085$ & \citet*{ch1-finley97} \\  
0102$-$185          &  72,370$\pm1793$ & $7.16\pm0.08$ & \citet{ch1-limoges10} \\
0556$-$375$^{\ast}$ &  72,800$\pm1800$ & $7.58\pm0.13$ & \citet{ch1-vennes97} \\  
1547$+$015$^{\ast}$ &  72,978$\pm2554$ & $7.628\pm0.111$ & \citet*{ch1-finley97} \\
1342$+$443$^{\ast}$ &  74,130$\pm2202$ & $7.84\pm0.11$ & \citet*{ch1-liebert05}$^{\dagger}$ \\ 
1201$-$049          &  74,798$\pm944$ & $7.475$ & \citet{ch1-voss06} \\
1312$-$253$^{\ast}$ &  75,463$\pm825$ & $7.682\pm0.027$ & \citet{ch1-voss06,ch1-koester09} \\
1547$+$015$^{\ast}$ &  75,585$\pm561$ & $7.612$ & \citet{ch1-voss06} \\
0158$-$227          &  75,758$\pm1128$ & $7.386$ & \citet{ch1-voss06} \\
0630$+$200          &  75,792$\pm751$ & $8.398\pm0.050$ & \citet*{ch1-finley97} \\  
1827$+$778          &  75,800$\pm610$ & $7.68\pm0.03$ & \citet{ch1-homeier98} \\
0616$-$084          &  76,320$\pm200$ & $8.05\pm0.15$ & \citet{ch1-vennes99} \\  
0111$-$381          &  76,857$\pm746$ & $7.367$ & \citet{ch1-voss06} \\
1749$+$717          &  76,900$\pm550$ & $7.56\pm0.03$ & \citet{ch1-homeier98} \\
1622$+$323          &  77,166$\pm1839$ & $7.838\pm0.082$ & \citet*{ch1-finley97} \\
0441$+$467$^{\ast}$ &  77,300$\pm3400$ & $7.31\pm0.14$ & \citet{ch1-bergeron94} \\  
0939$+$262          &  77,300$\pm1400$ & $7.78\pm0.06$ & \citet{ch1-bergeron94} \\  
0229$-$481          &  77,421$\pm2550$ & $7.549\pm0.064$ & \citet*{ch1-brag95} \\
1547$+$015$^{\ast}$ &  78,101$\pm3409$ & $7.49$ & \citet*{ch1-liebert05}$^{\dagger}$ \\ 
1342$+$443$^{\ast}$ &  78,700$\pm2700$ & $7.82\pm0.11$ & \citet{ch1-bergeron94} \\
1253$+$378$^{\ast}$ &  79,900$\pm3600$ & $6.61\pm0.20$ & \citet{ch1-bergeron94} \\  
1253$+$378$^{\ast}$ &  79,900$\pm3765$ & $6.61\pm0.21$ & \citet*{ch1-liebert05} \\  
2146$-$433          &  81,638$\pm1245$ & $7.994$ & \citet{ch1-voss06} \\
0441$+$467$^{\ast}$ &  83,800$\pm1700$ & $7.17\pm0.13$ & \citet{ch1-bergeron94} \\  
1305$-$017          &  85,773$\pm992$ & $7.800$ & \citet{ch1-voss06} \\
0345$+$006          &  86,850$\pm3544$ & $7.08\pm0.12$ & \citet{ch1-limoges10} \\
1738$+$669$^{\ast}$ &  88,010$^{+2390}_{-2610}$ & $7.79^{+0.14}_{-0.09}$ & \citet{ch1-marsh97} \\  
0500$-$156          &  94,488$\pm112$ & $7.214$ & \citet{ch1-voss06} \\
1738$+$669$^{\ast}$ &  95,324$\pm1217$ & $7.864\pm0.035$ & \citet*{ch1-finley97} \\  
0615$+$655          &  98,000$\pm5500$ & $7.07\pm0.15$ & \citet{ch1-homeier98} \\
2246$+$066          &  98,000$\pm1700$ & $7.04\pm0.05$ & \citet{ch1-homeier98} \\
0950$+$139          & 108,390$\pm$16,786 & $7.39\pm0.0.38$ & \citet*{ch1-liebert05} \\  
0948$+$534 & 136,762$\pm9805$ & $7.222$ & \citet*{ch1-liebert05}$^{\dagger}$ 
\enddata
\tablenotetext{a}{This list was compiled from the literature with input and advice from J.\ Holberg (private communication) to select the most reliable temperature estimates obtained from the highest quality optical and far-ultraviolet spectra.}
\tablenotetext{\ast}{Duplicate listing with independent parameter estimates.}
\tablenotetext{\dagger}{Revised parameters provided by J.\ Holberg (private communication).}

\end{deluxetable}

\begin{deluxetable}{l c c}
\tablecolumns{3} 
\tablewidth{0pt}
\tablecaption{The Hottest DAO White Dwarfs\tablenotemark{a} \label{t:ch1-hotdao}}
\tablehead{
\colhead{ } & 
\multicolumn{2}{c}{Effective Temperature (K)} \\
\multicolumn{1}{l}{Name} & 
\colhead{Balmer Lines} & 
\colhead{Lyman Lines}  
}
\startdata 
Abell 7       & 66,955 & \phn99,227 \\
HS\,0505+0112 & 63,227 &    120,000 \\
PuWe 1        & 74,218 &    109,150 \\
RE\,0720-318  & 54,011 & \phn54,060 \\
Ton 320       & 63,735 & \phn99,007 \\
PG\,0834+500  & 56,470 &    120,000 \\
Abell 31      & 74,726 & \phn93,887 \\
HS\,1136+6646 & 61,787 &    120,000 \\
Feige 55      & 53,948 & \phn77,514 \\
PG\,1210+533  & 46,338 & \phn46,226 \\
LB2           & 60,294 & \phn87,662 \\
HZ\,34        & 75,693 & \phn87,004 \\
Abell 39      & 72,451 & \phn87,965 \\
RE\,2013+400  & 47,610 & \phn50,487 \\
DeHt5         & 57,493 & \phn59,851 \\
GD\,561       & 64,354 & \phn75,627
\enddata
\tablenotetext{a}{Data used in this table are from \citet{ch1-good04}.}
\end{deluxetable}

\begin{deluxetable}{l c c c c c c c}
\tablecolumns{3} 
\tablewidth{0pt}
\tablecaption{DO White Dwarf Parameters\tablenotemark{a} \label{t:ch1-do}}
\tablehead{
\colhead{ } &
\colhead{ } &
\colhead{ } &
\multicolumn{5}{c}{Metal Abundances\tablenotemark{b} } \\
\multicolumn{1}{l}{Star} & 
\colhead{$T_{\rm eff}$ (kK)} &
\colhead{$\log{g}$} & 
\colhead{C} & 
\colhead{N} & 
\colhead{O} & 
\colhead{Fe} & 
\colhead{Ni} 
}
\startdata 
PG\,1034$+$001  &    100 & 7.5 & \phm{$<$ }$-5.0$  & \phm{$<$ }$-3.2$  & \phm{$<$ }$-4.1$  & \phm{$<$ }$-5.0$  &           $<-5.0$ \\
PG\,0108$+$101  & \phn95 & 7.5 & \phm{$<$ }$-2.0$  &           $<-7.3$ & \phm{$<$ }$-3.3$  & \phm{$<$ }$-4.3$  & \phm{$<$ }$-4.3$ \\
RE\,0503$-$289  & \phn70 & 7.5 & \phm{$<$ }$-2.3$  & \phm{$<$ }$-4.8$  & \phm{$<$ }$-3.3$  &           $<-6.0$ & \phm{$<$ }$-5.0$ \\
HS\,0111$+$0012 & \phn65 & 7.8 & \phm{$<$ }$-3.0$  &           $<-8.5$ &           \nodata &           $<-6.0$ &           $<-6.0$ \\
HZ\,21          & \phn53 & 7.8 &           $<-6.0$ & \phm{$<$ }$-5.0$  &           $<-6.0$ &           \nodata &           \nodata \\
HD\,149499\,B   & \phn50 & 8.0 &           $<-6.0$ &           $<-6.0$ &           $<-6.0$ &           \nodata &           \nodata
\enddata
\tablenotetext{a}{Data used in this table are from \citet{ch1-dreizler99}.}
\tablenotetext{b}{{Logarithm} of number ratios relative to He from homogeneous model atmosphere fits.}
\end{deluxetable}

\clearpage

\clearpage

\clearpage

\clearpage

\clearpage

\end{document}